\documentclass[11pt,a4paper,twocolumn]{article}

\usepackage[T1]{fontenc}
\usepackage[utf8]{inputenc}
\usepackage{microtype}
\usepackage{lmodern}
\usepackage[margin=1.8cm]{geometry}
\usepackage{amsmath,amssymb,amsfonts}
\usepackage{graphicx}
\usepackage{booktabs}
\usepackage{array}
\usepackage{tabularx}
\usepackage{xcolor}
\usepackage[hidelinks]{hyperref}
\usepackage{listings}
\usepackage{parskip}
\usepackage{caption}
\usepackage{pmboxdraw} 

\title{SHroom: A Python Framework for Ambisonics Room Acoustics Simulation and Binaural Rendering}
\author{Yhonatan Gayer, Boaz Rafaely\\[0.4em]\small Ben Gurion University of the Negev}
\date{}
\begin{document}
\maketitle

\begin{abstract}
Spatial audio research for virtual and augmented reality, teleconferencing and hearing devices often represents sound fields in the Spherical Harmonics (SH) domain, known as Ambisonics. A typical study simulates a room, renders what a listener or a microphone array would capture in it, and processes those signals in the SH domain. We present SHroom (Spherical Harmonics ROOM), an open-source Python library that performs this whole workflow in one package, from room simulation to binaural rendering, head rotation, microphone-array simulation and Ambisonics encoding.

Existing tools cover either the room simulation or the downstream processing, so researchers bridge them with ad-hoc code that is hard to reproduce and compare. In SHroom, every step operates on one shared signal type through one processing interface, so a simulated room flows through the complete chain without format conversions. Built on the image-source engine of \texttt{pyroomacoustics}, SHroom reproduces the Ambisonic Room Impulse Response (ARIR) of its spherical-harmonic receivers while computing it about $3\times$ faster for SH orders 4 to 12. SHroom is available at \href{https://github.com/Yhonatangayer/shroom}{https://github.com/Yhonatangayer/shroom} and installable via \texttt{pip install pyshroom}.
\end{abstract}

\section{Statement of Need}
\label{sec:need}

The development and evaluation of spatial audio algorithms based on Ambisonics rely on the ability to simulate acoustic environments, process Ambisonics and other types of signals, and render binaural or loudspeaker signals. A typical research workflow encompasses various and diverse computational components: (i) the generation of Ambisonic Room Impulse Responses (ARIRs), (ii) the generation of Ambisonics signals by convolving dry audio signals with the ARIRs, (iii) Spherical Harmonics (SH) domain filtering, such as convolution with Head-Related Transfer Functions (HRTFs) or microphone array Acoustic Transfer functions (ATFs) to produce signals at the ears or at the microphones, (iv) additional SH-domain processing including operations such as Wigner-D based listener head rotations, modeling of spherical microphone array prototypes, processing of SH-domain HRTFs, and encoding of Ambisonics from microphones arrays.

Implementing all these computational components within a single computation tool would streamline the research workflow. However, existing software can only provide implementations for various parts of such workflow.

\section{State of the Field}
\label{sec:sota}

For stages (i) and (ii), several existing room-acoustics simulators can generate ARIRs and convolve them with dry audio signals to produce Ambisonic signals. As summarized in Table~\ref{tab:sota}, \texttt{pyroomacoustics} (PRA)~\cite{pyroomacoustics-Scheibler2018}, MASP~\cite{perezlopez2020AES}, shoebox-roomsim~\cite{politis2016roomsim}, MCRoomSim~\cite{wabnitz2010ISRA}, and SAF~\cite{saf2024framework} generate ARIRs using the image-source method (ISM), whereas GSound-SIR~\cite{zang2025GSoundSIR} employs ray tracing (RT). These tools therefore cover room simulation and Ambisonics signal generation, but differ in the downstream processing they offer, as detailed in (iii) and (iv) above.

For stage (iii), SH-domain filtering, such as convolution with HRTFs or ATFs support is fragmented (see column `SH domain ATF processing' in Table~\ref{tab:sota}). Tools such as MASP~\cite{perezlopez2020AES}, SAF~\cite{saf2024framework}, \texttt{spaudiopy}~\cite{hold2025spaudiopy}, \texttt{sound-field-analysis-py}, and \texttt{pyfar}/\texttt{spharpy} provide dedicated SH-domain ATF processing. In contrast, room simulators like \texttt{shoebox-roomsim}~\cite{politis2016roomsim} and \texttt{MCRoomSim}~\cite{wabnitz2010ISRA} do not provide post-simulation SH-domain ATF filtering.

Stage (iv) encompasses additional processing such as HRTF pre-processing (e.g.\ MagLS equalization), the application of SH rotation matrices for listener head tracking, modeling of open and rigid spherical microphone array prototypes, and Ambisonics array encoding. Specialized toolboxes such as \texttt{spaudiopy}~\cite{hold2025spaudiopy} provide these capabilities, but omit room-acoustic engines altogether. Other packages cover only a subset of these features; for example, \texttt{pyfar}/\texttt{spharpy} lacks MagLS optimization, MASP~\cite{perezlopez2020AES} and \texttt{sound-field-analysis-py} do not include native rotation matrices, and room simulators (\texttt{pyroomacoustics}, \texttt{GSound-SIR}, \texttt{shoebox-roomsim}, \texttt{MCRoomSim}) bypass stage (iv) entirely. While SAF~\cite{saf2024framework} supports a comprehensive feature set, it does so through standalone C/C++ modules rather than a unified, scriptable workflow. In conclusion, no toolbox is currently available which offers this wide range of computations, which may be important in spatial audio processing research.

\paragraph{Contributions.} To address the fragmented availability of tools within this spatial audio workflow, this paper introduces \mbox{SHroom}, an open-source Python library designed to unify room acoustics and SH processing. The main contributions of this work are:
\begin{itemize}
    \item \textbf{An end-to-end SH-domain pipeline:} \mbox{SHroom} bridges the gap between room-acoustic simulation (stages i--ii) and downstream spatial processing (stages iii--iv). It is the first Python framework that allows users to generate ARIRs and seamlessly use them in advanced SH-domain processing without relying on external wrappers.
    \item \textbf{Comprehensive stage (iv) capabilities:} \mbox{SHroom} natively integrates SH processing. Each module implements an established method, all of which operate on the same ARIR and compose seamlessly with one another. Capabilities include Wigner-D rotation matrices, simulation of prototype and arbitrary microphone arrays, robust spatial encoding from arbitrary array geometries via the Ambisonics Signal Matching (ASM)~\cite{ASM,Parametric-ASM-like-paper} and Binaural Signal Matching (BSM)~\cite{BSM_journal_paper,Shai-paper}, and HRTF preprocessing via MagLS~\cite{HRTF_MagLS,kassakian2006convex,Ambisonics_MagLS} and array-aware MagLS HRTF (AA-MagLS)~\cite{gayer2026TASLP}.
\end{itemize}

\begin{table*}[tp]
\centering\footnotesize
\setlength{\tabcolsep}{2pt}
\begin{tabular}{>{\raggedright\arraybackslash}p{3.5cm}p{2.45cm}>{\centering\arraybackslash}p{2.35cm}>{\centering\arraybackslash}p{2.25cm}>{\centering\arraybackslash}p{1.35cm}>{\centering\arraybackslash}p{1.15cm}>{\centering\arraybackslash}p{2.00cm}>{\centering\arraybackslash}p{1.45cm}}
\toprule
Software & Language & Room sim. \& ARIR & SH domain ATF processing & MagLS & SH rotation & SH array simulation & Array encoding\\
\midrule
pyroomacoustics & Python & \checkmark (ISM) & - & - & - & - & - \\
GSound-SIR & Python & \checkmark (RT) & - & - & - & - & - \\
MASP & Python & \checkmark (ISM) & \checkmark & - & - & \checkmark & \checkmark \\
shoebox-roomsim & MATLAB & \checkmark (ISM) & - & - & - & - & - \\
MCRoomSim & MATLAB & \checkmark (ISM) & - & - & - & - & - \\
SAF & C / C++ & \checkmark (ISM) & \checkmark & \checkmark & \checkmark & \checkmark & \checkmark \\
spaudiopy & Python & - & \checkmark & \checkmark & \checkmark & \checkmark & \checkmark \\
sound-field-analysis-py & Python & - & \checkmark & - & - & \checkmark & \checkmark \\
pyfar / spharpy & Python & - & \checkmark & - & \checkmark & \checkmark & \checkmark \\
\midrule
\textbf{SHroom} & \textbf{Python} & \textbf{\checkmark~(ISM)} & \textbf{\checkmark} & \textbf{\checkmark} & \textbf{\checkmark} & \textbf{\checkmark} & \textbf{\checkmark} \\
\bottomrule
\end{tabular}
\caption{Capability comparison of open-source spatial-audio software. \checkmark~= provided as a first-class feature; -~= not provided. ``Language'' refers to the user-facing interface language of the software. ``Room sim.\ \& ARIR'' means the simulated room field is exposed to the user as SH coefficients. ``SH array simulation'' means modal simulation of rigid or open spherical arrays with radial filters, not merely the placement of pressure microphones in a room. ``Array encoding'' means encoding those signals from arbitrary array geometries. SAF provides the individual capabilities as independent C modules rather than as one composable pipeline.}
\label{tab:sota}
\end{table*}

\section{Software Design}
\label{sec:design}

SHroom is organised around one data type and one interface. \texttt{SpatialSignal} holds a multichannel signal together with its sampling rate, spatial grid and two domain flags (time or frequency; space or SH), and records every transform applied to it. Domain changes (\texttt{toFreq}, \texttt{toSH}, \texttt{toSpace}) and rotations (Wigner-D in the SH domain, or grid rotation in the space domain) are methods on this type, so every module accepts and returns the same object and checks the domain it needs rather than assuming it.

Processing steps (\texttt{BinauralDecoder}, \texttt{ArrayDecoder}, \texttt{ASMEncoder}, \texttt{BSMEncoder}) implement a single \texttt{process} method from \texttt{SpatialSignal} to \texttt{SpatialSignal}. They can therefore be applied in any order that is physically meaningful, and a \texttt{ProcessorChain} collapses a sequence of them into one equivalent filter by passing unit impulses through the chain once, so a long signal is convolved only a single time.

The room model deliberately reuses the \texttt{pyroomacoustics} image-source engine for geometry, wall materials and reflection bookkeeping instead of re-implementing it, and replaces only the receiver stage: all image sources are projected onto the SH basis in one vectorised evaluation and placed with the same fractional-delay filters, rather than simulating one directional microphone per SH channel. Frequency-dependent absorption is handled per octave band, as in \texttt{pyroomacoustics}. The trade-offs are explicit: SHroom supports shoebox rooms and the image-source method only, and it rejects \texttt{ray\_tracing=True} instead of silently ignoring it; randomised image-source positions are enabled by default to reduce the metallic sound of the sparse late field, and \texttt{Room.ism\_coverage()} warns when the reflection order is too low for the room's reverberation time.

\section{Research Impact Statement}
\label{sec:impact}

SHroom provides open, tested Python implementations of ASM~\cite{ASM}, BSM~\cite{BSM_journal_paper}, MagLS~\cite{HRTF_MagLS} and AA-MagLS~\cite{gayer2026TASLP}. Its correctness is checked against independent references: driven from the same image-source engine, its ARIR reproduces the one obtained from \texttt{pyroomacoustics}' spherical-harmonic receivers to float32 precision (relative error $6 \times 10^{-8}$) while building it about $3\times$ faster for $4 \le N \le 12$, and its BSM encoder is validated against a MATLAB reference. The repository ships benchmark scripts for this comparison and for the convergence of the ASM, spectrally equalized ASM, BSM and AA-MagLS encoders with SH order, and six runnable examples covering each binaural rendering path (HRTF, MagLS, ASM, ASM with AA-MagLS, BSM, and head rotation).

SHroom is released on PyPI (\texttt{pip install pyshroom}), where it has been downloaded about 1,000 times since its first release in March 2026 (pypistats, excluding mirrors). It is tested by a 212-test suite in continuous integration across Python 3.9--3.13, and ships contribution guidelines and citation metadata.

\section{AI Usage Disclosure}
\label{sec:ai}

Generative AI tools were used to assist with software documentation, code refactoring, and the preparation of this paper (including editing for clarity and structure). All AI-suggested content was reviewed by the authors, and correctness was verified through the project's automated test suite and continuous-integration pipeline together with manual inspection; the algorithms, design decisions, and reported results are the authors' own. No generative AI was used to produce the experimental results.

\section*{Acknowledgements}

The bundled HRTF data is the Neumann KU 100 spherical far-field HRIR compilation~\cite{HRTF_data_set}. Room geometry and image-source computation build on \texttt{pyroomacoustics}~\cite{pyroomacoustics-Scheibler2018}, and the MagLS implementation follows the formulation of Sch\"orkhuber et al.~\cite{HRTF_MagLS}.

\bibliographystyle{IEEEtran}
\bibliography{ref}

@inproceedings{Shai-paper,
  author    = {Hermon, Shai and Tourbabin, Vladimir and Ben-Hur, Zamir and Donley, Jacob and Rafaely, Boaz},
  title     = {{Binaural signal matching with an arbitrary array based on a sound field model}},
  booktitle = {Proceedings of the AES International Conference on Audio for Virtual and Augmented Reality (AVAR)},
  month     = {August},
  year      = {2022},
  address   = {Redmond, WA, USA},
  publisher = {Audio Engineering Society},
  url       = {https://www.aes.org/e-lib/browse.cfm?elib=21861}
}

@article{Parametric-ASM-like-paper,
  author  = {McCormack, Leo and Politis, Archontis and Gonzalez, Raimundo and Lokki, Tapio and Pulkki, Ville},
  journal = {IEEE/ACM Transactions on Audio, Speech, and Language Processing},
  title   = {{Parametric Ambisonic Encoding of Arbitrary Microphone Arrays}},
  year    = {2022},
  volume  = {30},
  pages   = {2062--2075},
  doi     = {10.1109/TASLP.2022.3182857}
}

@misc{HRTF_data_set,
  author    = {Bernsch{\"u}tz, Benjamin},
  title     = {{Spherical Far-Field HRIR Compilation of the Neumann KU100}},
  year      = {2020},
  publisher = {Zenodo},
  month     = {7},
  doi       = {10.5281/zenodo.3928297},
  url       = {https://doi.org/10.5281/zenodo.3928297}
}

@inproceedings{HRTF_MagLS,
  title     = {{Binaural rendering of ambisonic signals via magnitude least squares}},
  author    = {Sch{\"o}rkhuber, Christian and Zaunschirm, Markus and H{\"o}ldrich, Robert},
  booktitle = {Proceedings of the DAGA},
  volume    = {44},
  pages     = {339--342},
  year      = {2018}
}

@article{Ambisonics_MagLS,
  author  = {L{\"u}beck, Tim and Helmholz, Hannes and Arend, Johannes and P{\"o}rschmann, Christoph and Ahrens, Jens},
  year    = {2020},
  month   = {07},
  pages   = {428--440},
  title   = {Perceptual Evaluation of Mitigation Approaches of Impairments due to Spatial Undersampling in Binaural Rendering of Spherical Microphone Array Data},
  volume  = {68},
  journal = {Journal of the Audio Engineering Society},
  doi     = {10.17743/jaes.2020.0038}
}

@inproceedings{ASM,
  author    = {Gayer, Yhonatan and Tourbabin, Vladimir and Ben-Hur, Zamir and Donley, Jacob and Rafaely, Boaz},
  booktitle = {2024 IEEE International Conference on Acoustics, Speech, and Signal Processing Workshops (ICASSPW)},
  title     = {Ambisonics Encoding For Arbitrary Microphone Arrays Incorporating Residual Channels For Binaural Reproduction},
  year      = {2024},
  pages     = {244--248},
  doi       = {10.1109/ICASSPW62465.2024.10627373}
}

@article{BSM_journal_paper,
  title   = {Design and Analysis of binaural signal matching with arbitrary microphone arrays},
  author  = {Madmoni, Lior and Ben-Hur, Zamir and Donley, Jacob and Tourbabin, Vladimir and Rafaely, Boaz},
  journal = {arXiv preprint arXiv:2408.03581},
  year    = {2024}
}

@phdthesis{kassakian2006convex,
  author  = {Kassakian, P. W.},
  title   = {Convex approximation and optimization with applications in magnitude filter design and radiation pattern synthesis},
  school  = {University of California, Berkeley},
  address = {Berkeley, CA},
  year    = {2006}
}

@inproceedings{pyroomacoustics-Scheibler2018,
  author    = {Scheibler, Robin and Bezzam, Eric and Dokmani{\'c}, Ivan},
  title     = {Pyroomacoustics: A Python Package for Audio Room Simulation and Array Processing Algorithms},
  booktitle = {IEEE International Conference on Acoustics, Speech and Signal Processing (ICASSP)},
  year      = {2018},
  pages     = {351--355},
  doi       = {10.1109/ICASSP.2018.8461310}
}

@article{gayer2026TASLP,
  title     = {Array-Aware Ambisonics and {HRTF} Encoding for Binaural Reproduction With Wearable Arrays},
  author    = {Gayer, Yhonatan and Tourbabin, Vladimir and Ben-Hur, Zamir and Lou-Alon, David and Rafaely, Boaz},
  journal   = {IEEE Transactions on Audio, Speech and Language Processing},
  year      = {2026},
  publisher = {IEEE}
}

@misc{hold2025spaudiopy,
  title        = {{spaudiopy}: Spatial Audio Python Package},
  author       = {Hold, Christoph},
  year         = {2025},
  note         = {Version 0.2.0, MIT licence},
  url = {https://github.com/chris-hld/spaudiopy}
}

@misc{saf2024framework,
  title        = {{Spatial\_Audio\_Framework}: A Cross-Platform Framework for Developing Spatial Audio Algorithms and Software in {C}/{C}++},
  author       = {McCormack, Leo and Politis, Archontis},
  year         = {2024},
  url = {https://github.com/leomccormack/Spatial_Audio_Framework}
}

@inproceedings{wabnitz2010ISRA,
  title     = {Room Acoustics Simulation for Multichannel Microphone Arrays},
  author    = {Wabnitz, Andrew and Epain, Nicolas and Jin, Craig and van Schaik, Andr{\'e}},
  booktitle = {Proceedings of the International Symposium on Room Acoustics (ISRA)},
  address   = {Melbourne, Australia},
  year      = {2010}
}

@misc{zang2025GSoundSIR,
  title         = {{GSound-SIR}: A Spatial Impulse Response Ray-Tracing and High-Order Ambisonic Auralization Python Toolkit},
  author        = {Zang, Yongyi and Kong, Qiuqiang},
  year          = {2025},
  eprint        = {2503.17866},
  archiveprefix = {arXiv}
}

@misc{politis2016roomsim,
  title        = {Shoebox Room Simulator for Spatial Sound Processing},
  author       = {Politis, Archontis},
  year         = {2016},
  url = {https://github.com/polarch/shoebox-roomsim},
  note         = {MATLAB/Octave library}
}

@inproceedings{perezlopez2020AES,
  title     = {A {Python} Library for Multichannel Acoustic Signal Processing},
  author    = {Perez-Lopez, Andres and Politis, Archontis},
  booktitle = {148th Convention of the Audio Engineering Society},
  note      = {Paper 569},
  year      = {2020}
}

\end{document}